\documentclass[amsmath, amsfonts, preprint, superscriptaddress, prl]{revtex4}
\usepackage{graphicx}
\usepackage{epsfig}
\usepackage{bm}
\usepackage{dcolumn}
\usepackage{amsmath}
\usepackage{amssymb}
\usepackage{color}
\usepackage{soul}
\begin{document}

\title{Superconductivity-localization interplay and fluctuation magnetoresistance in epitaxial BaPb$_{1-x}$Bi$_x$O$_3$ thin films}

\author{D.T. Harris}
\affiliation{Department of Materials Science and Engineering, University of Wisconsin-Madison, Madison, WI 53706, USA}
\author{N. Campbell}
\affiliation{Department of Physics, University of Wisconsin-Madison, Madison, WI 53706, USA}
\author{R. Uecker}
\affiliation{Leibniz Institute for Crystal Growth, Berlin, Germany}
\author{M. Br{\"u}tzam}
\affiliation{Leibniz Institute for Crystal Growth, Berlin, Germany}
\author{D.G. Schlom}
\affiliation{Department of Materials Science and Engineering, Cornell University, Ithaca, NY 14853, USA}
\affiliation{Kavli Institute at Cornell for Nanoscale Science, Ithaca, NY 14853, USA}
\author{A. Levchenko}
\affiliation{Department of Physics, University of Wisconsin-Madison, Madison, WI 53706, USA}
\author{M.S. Rzchowski}
\affiliation{Department of Physics, University of Wisconsin-Madison, Madison, WI 53706, USA}
\author{C.-B. Eom}
\email{eom@engr.wisc.edu}
\affiliation{Department of Materials Science and Engineering, University of Wisconsin-Madison, Madison, WI 53706, USA}

\date{\today}

\begin{abstract} 
BaPb$_{1-x}$Bi$_x$O$_3$ is a superconductor, with transition temperature $T_c=11$ K, whose parent compound BaBiO$_3$ possess a charge ordering phase and perovskite crystal structure reminiscent of the cuprates.  The lack of magnetism simplifies the BaPb$_{1-x}$Bi$_{x}$O$_3$ phase diagram, making this system an ideal platform for contrasting high-$T_c$ systems with isotropic superconductors. Here we use high-quality epitaxial thin films and magnetotransport to demonstrate superconducting fluctuations that extend well beyond $T_c$. For the thickest films (thickness above $\sim100$ nm) this region extends to $\sim27$ K, well above the bulk $T_c$ and remarkably close to the higher $T_c$ of Ba$_{1-x}$K$_x$BiO$_3$ ($T_c=31$ K). We drive the system through a superconductor-insulator transition by decreasing thickness and find the observed $T_c$ correlates strongly with disorder. This material manifests strong fluctuations across a wide range of thicknesses, temperatures, and disorder presenting new opportunities for understanding the precursor of superconductivity near the 2D-3D dimensionality crossover.
\end{abstract}

\maketitle

In contrast to the layered cuprate superconductors, BaPb$_{1-x}$Bi$_x$O$_3$ (BPBO, $T_c=11$ K) and Ba$_{1-x}$K$_x$BiO$_3$ (BKBO, $T_c=31$ K) are isotropic and nonmagnetic, however, there are still interesting similarities~\cite{Sleight,Cava}. The bismuthates are complex oxides with oxygen octahedra similar to the cuprates, and the parent insulating BaBiO$_3$ (BBO) possesses a competing phase, a charge density wave (CDW), which is suppressed for superconducting compositions. The study of the simpler, conventional bismuthate may lead to a deeper understanding of the role of CDW physics in the more complicated cuprates. The cuprate phase diagram is characterized by numerous electronic and magnetic phases and the properties are strongly influenced by disorder \cite{Phillips}. In thin conventional superconductors, disorder can lead to a pseudogap reminiscent of the high-$T_c$ cuprates, suggesting a possible connection between the layered cuprate structure and dimensionally confined conventional superconductors~\cite{Sacepe}. In superconducting BPBO single crystals, Luna \textit{et al.}~\cite{Luna} found a reduction in the density of states consistent with a disorder-driven metal-insulator transition and predicted a disorder-free $T_c$ of 17 K in the strong coupling limit and 52 K in the weak coupling limit for $x=0.25$.
 
Here we demonstrate an extended region of positive magnetoresistance in epitaxial thin films of BaPb$_{0.75}$Bi$_{0.25}$O$_3$ that is well described by superconducting fluctuations. This fluctuation regime persists for the thickest films that are well within the 3D regime, consistent with the high disorder found in our films. Restricting film thickness causes a superconductor-to-insulator transition (SIT) that correlates with disorder. Although our results are consistent with the disorder levels found in bulk single crystals~\cite{Luna}, we find that the critical thickness for superconductivity depends on extrinsic factors related to the poor lattice matching of BPBO with common perovskite substrates.
 
\begin{figure}
		\centering
		\includegraphics[width=16cm]{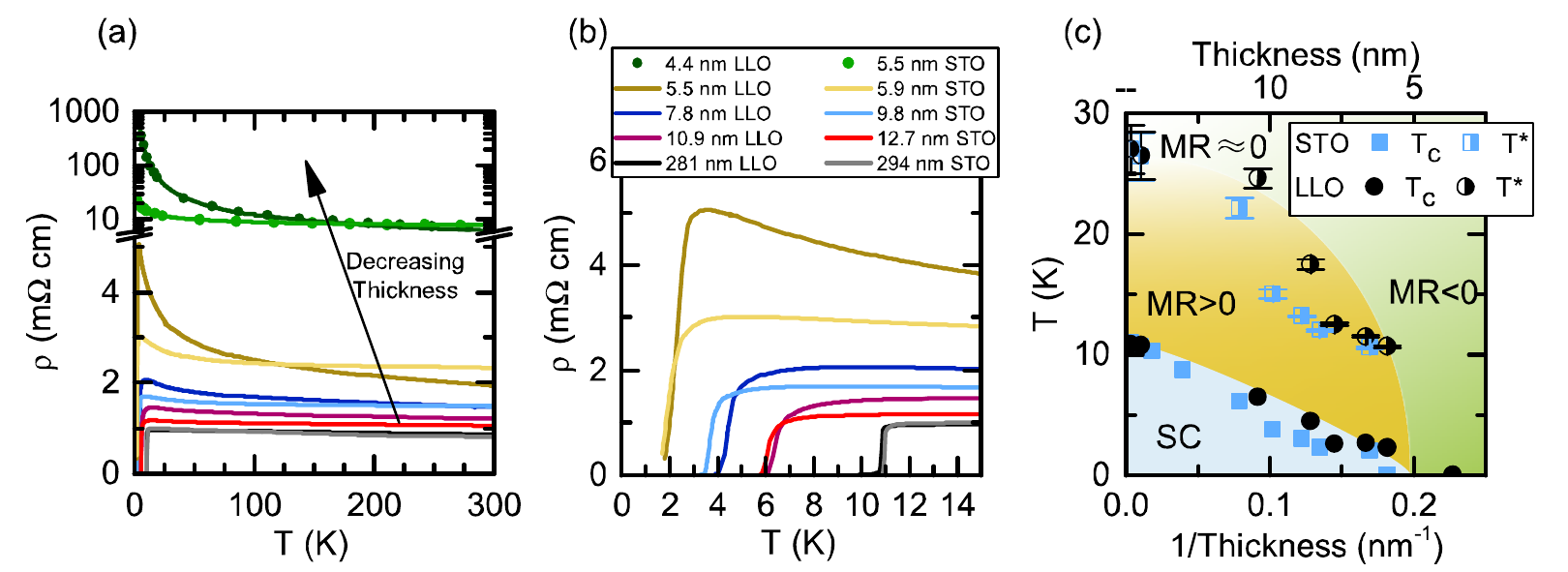}
		\caption{Resistivity measurements from 300 K to 2 K. (a) Representative resistivity measurements for films on LLO and STO. For 4.4 nm on LLO and 5.5 nm on STO, the dots are experimental data and the solid line is a fit to a variable range hopping model. (b) Transition region for superconducting samples from (a). (c) Phase diagram of the BPBO system vs thickness for films on LLO (black circles) and STO (blue squares). The shaded regions are guides to the eye.}\label{Fig1}
\end{figure}
 
The high-quality epitaxial growth of BBO-based materials presents additional challenges since B(K,Pb)BO exhibits one of the largest lattice parameters ($a_{pc} = 4.26-4.36$ \AA) among the ABO$_3$ perovskites. There is little understanding of how the large lattice mismatch, $>10\%$ on typical commercial perovskite substrates such as SrTiO$_3$ (STO), impacts the structural and electronic properties of these materials. Reports in the literature demonstrate epitaxial growth on STO~\cite{Hellman1,Suzuki,Plumb}, MgO ($a = 4.21$ \AA)~\cite{Hellman2,Prieto,Inumaru}, or by using buffer layers~\cite{Lee}. We use recently developed LaLuO$_3$ (LLO, $a_{pc} = 4.187$ \AA) single crystals~\cite{Uecker} as a substrate to grow BPBO ($x=0.25, a_{pc} = 4.29$ \AA) and BBO films, demonstrating that reduction of the lattice mismatch from $\sim10.1\%$ (STO) to $\sim2.7\%$ (LLO) improves crystallinity, surface roughness, and superconducting transitions.

\begin{figure}
		\centering
		\includegraphics[width=16cm]{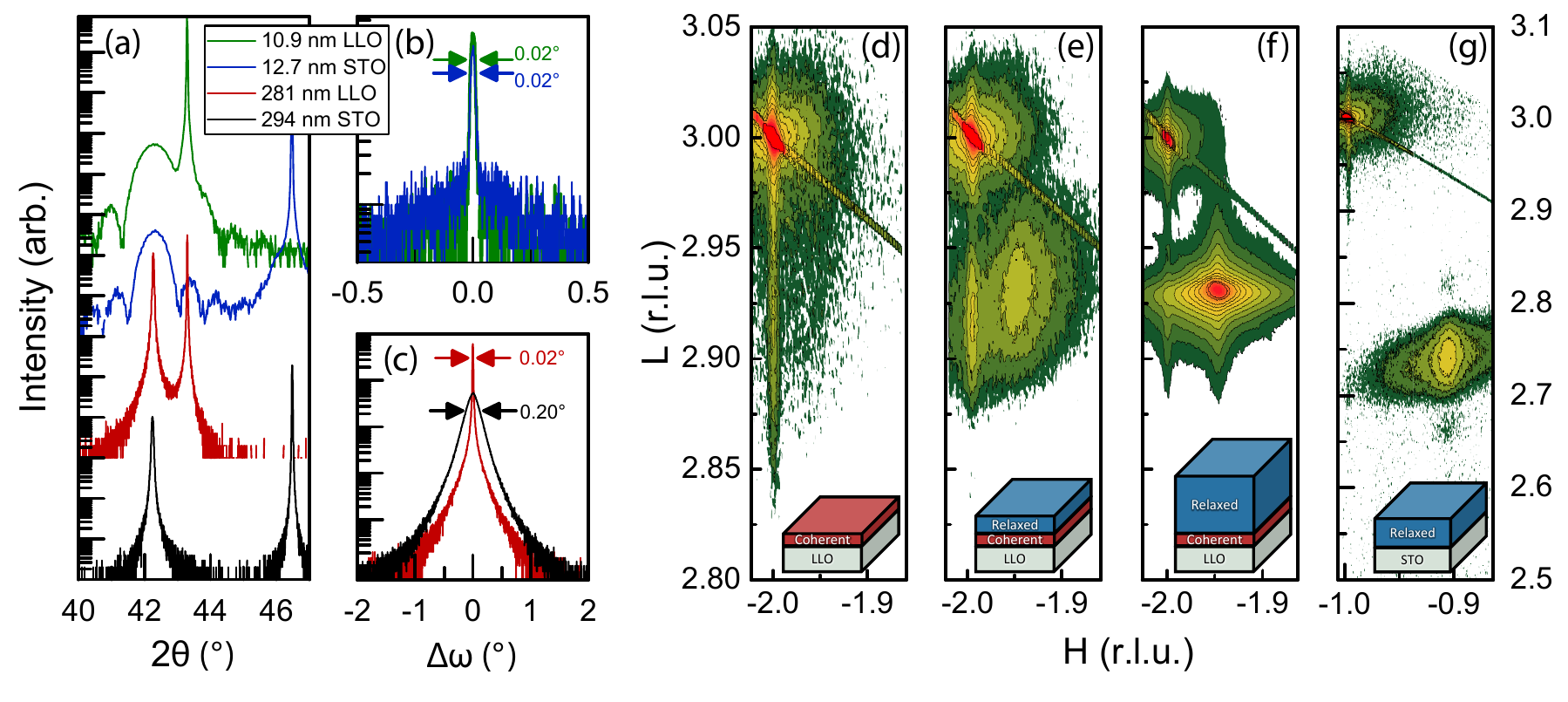}
		\caption{X-ray diffraction characterization. (a) Representative out of plane $2\theta-\omega$ scans around the 002$_{pc}$ peak for thin ($\sim$11 nm) and thick ($\sim$290 nm) BPBO films on STO and LLO. The corresponding rocking curves around the 002$_{pc}$ film peaks for (b) thin and (c) thick films with the full width at half maxes indicated. Reciprocal space maps around the 103$_{pc}$ reflection for (d) a fully coherent 4.4 nm thick film on LLO, as well bilayer relaxed films on LLO (e) 7.8 nm thick and (f) 281 nm thick and (g) fully relaxed 8.2 nm on STO. The inset cartoons show the film structure for each space map.}\label{Fig2}
\end{figure}

Epitaxial films of optimally doped $x=0.25$ BPBO were grown using $90^{\circ}$ off-axis rf-magnetron sputtering on (001) STO and (110) LLO substrates (see Supplemental Material for detailed methods). Thick films grown on both substrates exhibit a room temperature resistivity of $0.8$m$\Omega$-cm, the lowest reported for BPBO with $x=0.25$, and a slightly negative $d\rho /dT$ typical for optimally doped BPBO, as seen in Fig. \ref{Fig1}~\cite{Suzuki,Thanh,GG1}. The thick films exhibit sharp transitions with transition widths of $0.2$ K ($90\%-10\%$ of normal state) and $T_c$ ($50\%$ of normal state) of 10.9 K, comparable to bulk single crystal~\cite{GG2} and polycrystalline ceramics~\cite{Thanh}, and the highest reported for thin films~\cite{Suzuki}. We also find no evidence of inhomogeneous transitions as seen in bulk single crystals~\cite{GG2}. The quality of BPBO films is further demonstrated in Figs. \ref{Fig2}(a)-\ref{Fig2}(c) by the narrow $\omega$-rocking curves, presence of Kiessig fringes, and out-of-plane lattice constant of 4.276 \AA, in good agreement with the bulk value.

Decreasing the film thickness leads to higher resistivity, more negative $d\rho /dT$, depressed $T_c$, and broadened transitions as shown in Figs. \ref{Fig1}(a) and \ref{Fig1}(b). The thinnest films on both LLO and STO show insulating behavior well fit by a 2D variable range hopping model ($T_0=$1070 K and 7.5 K respectively) over the entire temperature range [solid lines in Fig. \ref{Fig1}(a)]. The LLO films exhibit higher $T_c$ when compared with films grown on STO. Fig. \ref{Fig1}(c) shows the extracted $T_c$ ($50\%$ of normal state), showing a smaller critical thickness for films on LLO ($d_c\sim5$ nm) than those on STO ($d_c\sim5.5$ nm).

Previous reports of BPBO polycrystalline films found significant suppression of $T_c$ already apparent at 200 nm, a much larger thickness than the onset of $T_c$ reduction in our epitaxial films~\cite{Hidaka}. Grain boundaries in polycrystalline BPBO form Josephson junctions~\cite{Takagi,Roshchin}. leading to reentrant behavior~\cite{Gantmakher}. Our epitaxial films do not show reentrant behavior, and the high crystalline quality evident in the rocking curves makes the existence of a granular structure unlikely.

\begin{figure}
		\centering
		\includegraphics[width=16cm]{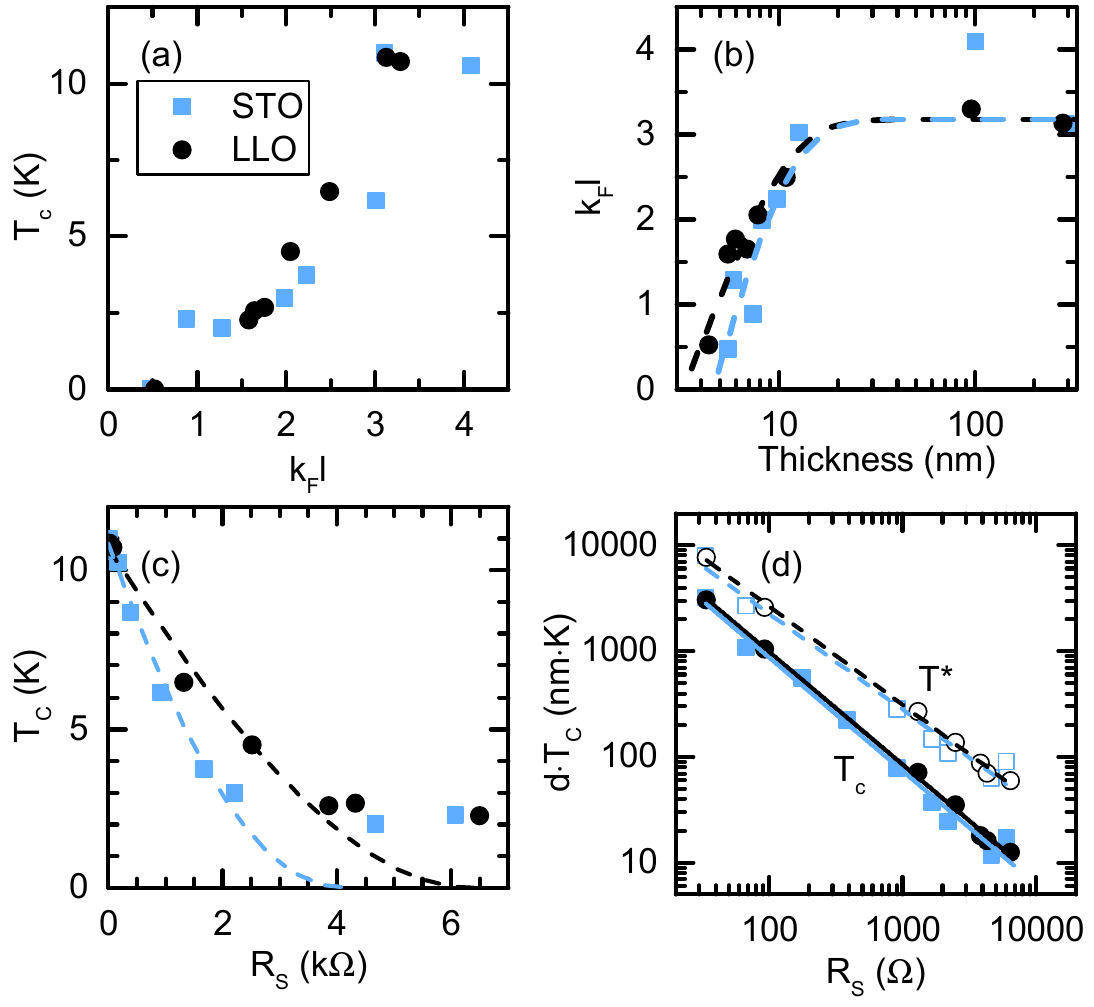}
		\caption{Disorder and superconductivity. (a) Transition temperature vs. the Mott-Ioffe-Regel parameter, $k_Fl$, (extracted from room temperature measurements), and (b) $k_Fl$ vs. thickness (the dashed lines are guides for the eye). (c) $T_c$ vs. the sheet resistance at 20 K. The dashed lines are fits to Finkel'stein's model for a homogenously disordered superconductor. (d) Power-law scaling dependence of thickness times transition temperature ($d \cdot T_c$) versus $R_S$ based on the phenomenological formula $d \cdot T_c=A R^{-B}_S$, where $A$ and $B$ are fitting parameters~\cite{Ivry}.}\label{Fig3}
\end{figure}

Disorder can be parameterized using the Mott-Ioffe-Regel parameter, $k_Fl$, determined from the normal state resistivity $\rho$ and the Hall coefficient $R_H$ by using the free electron formula $k_Fl=(3\pi^2)^{2/3}\hbar R^{1/3}_H/(\rho e^{5/3})$, which describes the limit of scattering in a system before localization occurs. In our system we find strong correlation between $k_Fl$ and $T_c$, as shown in Fig. \ref{Fig3}, with films on LLO showing higher $k_Fl$ for a given thickness. The disorder-induced superconductor-insulator transition, controlled by a variety of parameters including thickness, has been extensively studied in conventional and high-temperature systems, with recent efforts focusing on the quantum nature of the SIT, however, no complete theoretical understanding exists for the variety of phenomena experimentally observed ~\cite{Gantmakher,Goldman}.

For 2D homogeneously disordered films, Finkel'stein developed a model for $T_c$ suppression from Coulomb interactions assuming no change to the bulk electron gas properties with changes to $R_S$~\cite{Finkelstein}.  The data for both sets of films initially fit well to Finkel'stein's model, Fig. \ref{Fig3}(c), indicating an increase in the scattering time for films on LLO compared to STO, however, the model breaks down for films close to the critical thickness. Epitaxial films can experience thickness dependent strain relaxation, resulting in changes to material structure and properties with decreasing thickness. We investigate the strain state of our films using reciprocal space maps, shown in Fig. \ref{Fig2}(d)-\ref{Fig2}(g). On STO, BPBO grows relaxed, as reported by other groups~\cite{Kim,Meir}. On LLO, we obtain coherent growth for films up to $\sim$4.5 nm, with an abrupt relaxation at $\sim$4.5 nm that forms a layered structure [see inset schema Figs. \ref{Fig2}(d)-\ref{Fig2}(f)] that remains present for the thickest films studied. Similar relaxation behavior has been reported for other oxide epitaxial systems~\cite{Oh,Gao,Daumont,Yamahara} and occurs in undoped BaBiO$_3$ films as shown in supplemental Fig. \ref{Fig6}. The relaxed BPBO phase on both substrates has lattice constants of $a_{pc}=4.29$ \AA\, and $c_{pc}=4.28$ \AA, in good agreement with bulk values from powder diffraction~\cite{Climent-Pascual}. Films on both substrates show variations in the out-of-plane lattice parameter at low thicknesses, see supplemental Fig. \ref{Fig7}, indicating assumptions are likely violated for the Finkel'stein model. Both sets of films show a power law relationship between $d \cdot T_c$ and $R_S$, shown in Fig. \ref{Fig3}(d), an empirical observation found in many thin superconducting systems that is not well understand theoretically~\cite{Ivry}. 

\begin{figure}
		\centering
		\includegraphics[width=16cm]{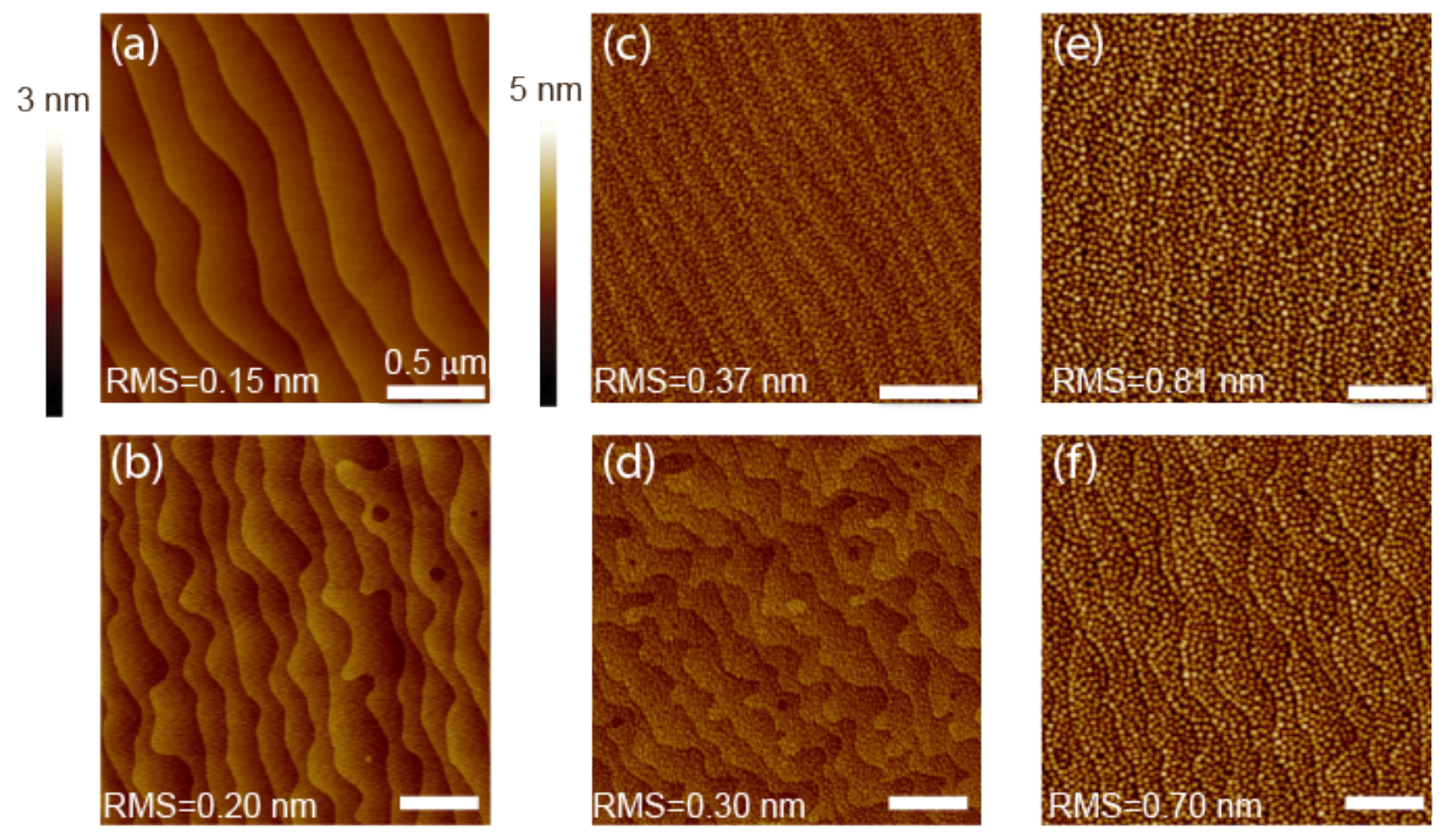}
		\caption{Atomic force microscopy images. Surface topography images for (a) treated STO and (b) LLO, (c) 12.7 nm BPBO on STO, (d) 10.9 nm on LLO, (e) 101 nm on STO, and (f) 96.5 nm on LLO.}\label{Fig4}
\end{figure}

The x-ray diffraction in Fig. \ref{Fig2} reveals rocking curves with higher diffuse backgrounds and RSMs with broader in-plane components for all films grown on STO. The mosaic spread seen in the broadening of rocking curves and reciprocal space maps is indicative of a high dislocation density~\cite{Biegalski,Darakchieva}. Improving the film-substrate lattice match by switching to LLO reduces the diffuse background, consistent with a reduction in dislocations and mosaicity. Additionally, surface and interface scattering become more important in thinner films and atomic force microscopy (AFM) consistently reveals smoother surfaces for films grown on LLO, as shown in Fig. \ref{Fig4}. The smaller critical thickness for superconductivity for films on LLO is consistent with the reduced disorder in the higher quality, smoother films~\cite{Ziman,Kramer,Timalsina}.
 
\begin{figure}
		\centering
		\includegraphics[width=16cm]{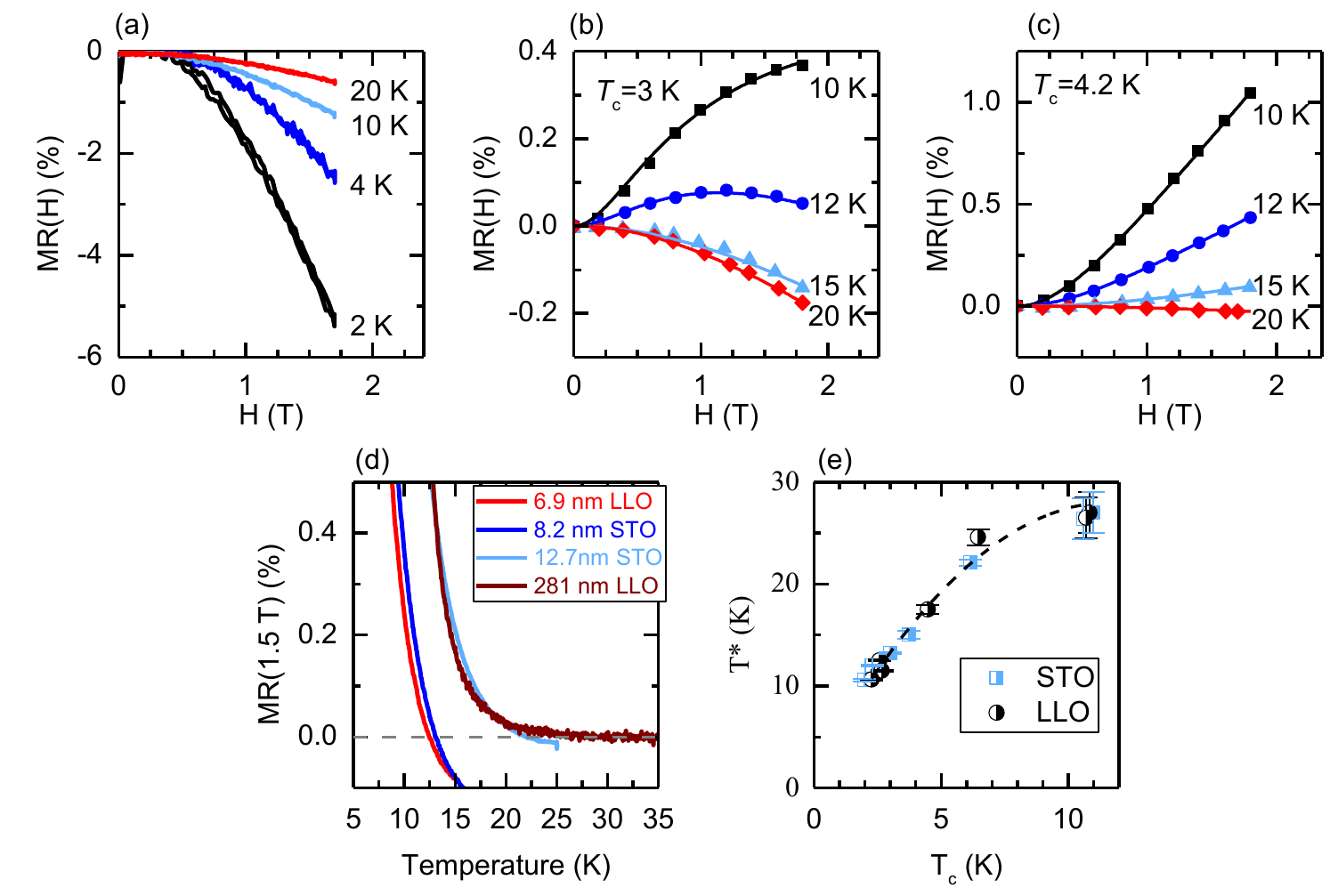}
		\caption{Magnetoresistance measurements above $T_c$. Magnetoresistance at fixed temperatures for (a) insulating 5.5 nm on STO, (b) superconducting 8.2 nm BPBO on STO, and (c) superconducting 7.8 nm BPBO on LLO. In (b) and (c) the markers are experimental data points with the solid lines fits to superconducting fluctuations and weak localization in a 2D disordered system. (d) Representative magnetoresistance vs temperature. (e) The MR inflection temperature $T^*$ vs $T_c$ (the dashed line is a guide for the eye).}\label{Fig5}
\end{figure}
 
Thin disordered superconductors routinely show evidence of superconductivity above the measured $T_c$~\cite{Larkin-Book,Gantmakher}. We investigated the insulating to superconducting transition via magnetotransport, revealing positive magnetoresistance (MR) in all superconducting films well above $T_c$. Non-superconducting films (with thickness less than a critical thickness $d_c \sim 5-6$ nm) show only negative magnetoresistance that increases in magnitude as the temperature is lowered, as shown in Fig. \ref{Fig5}(a) for an insulating film on STO. Magnetoresistance in the insulating phase of disordered superconductors can show a variety of responses. For instance, positive magnetoresistance reported in insulating samples is evidence for existence of localized superconductivity persisting beyond the superconductor-insulator transition~\cite{Baturina,Chand}. However, the observation of positive magnetoresistance in insulating samples is not universal even within the same material system, suggesting that localization can occur through different mechanisms of disorder resulting in distinct insulator states~\cite{Larkin-SIT,Stewart,Shahar-1,Shahar-2}.
  
In contrast to the insulating samples, superconducting films show an extended region of positive magnetoresistance above $T_c$ that extends to a temperature $T^*$ that is highly correlated with $T_c$, as shown in Fig. \ref{Fig5}(e), suggesting a strong link between the positive MR and superconductivity~\cite{Bergmann,Gordon,Shinozaki,Giannouri}. In superconductors, a positive magnetoresistance above $T_c$ is associated with the presence of superconducting fluctuations and can arise from several different mechanisms~\cite{Larkin-Book}. For thin $\sim$8 nm films, we fit our experimental MR data to models for fluctuations and weak localization in 2D disordered systems, see supplemental information, and find excellent agreement in the region far above the transition and at low fields, as shown in Figs. \ref{Fig5}(b) and \ref{Fig5}(c). We note that while we used the readily available expressions for magnetoresistance of a 2D film~\cite{Larkin-Book,Varlamov,Schwiete} and obtained excellent agreement with experiment, far from $T_c$ we expect a shortened coherence length and that our films are likely in the 3D regime~\cite{AL}. The theory of fluctuations for 3D systems is incomplete with no expressions for the field dependence for all terms available, in part due to difficulty in preparing metallic systems with high enough disorder for measurable fluctuations~\cite{Rosenbaum}. The BPBO system exhibits measurable fluctuations across a wide range of thicknesses and at temperatures easily accessed by common $He$ cryostats, allowing comparison with theory and study of the cross-over regime from 2D to 3D. Although we are beyond the strict limits of 2D fluctuation theory, the correlation between $T_c$ and $T^*$ [Fig. \ref{Fig5}(e)] and the good fits of experimental data strongly suggest the positive MR originates from superconducting fluctuations. Weak anti-localization (WAL) can also lead to positive MR in systems with strong spin-orbit coupling. Separation of the various terms that contribute to MR is difficult, but we point out the strong correlation between $T_c$ and $T^*$ [Fig. \ref{Fig5}(e)], the power-law scaling of $T^*$ in a manner similar to $T_c$ [Fig. \ref{Fig3}(d)], and the smooth divergence of MR as $T_c$ is approached [Fig. \ref{Fig5}(d)]. Further demonstration of the presence of fluctuations is beyond the scope of the present study but could be searched for in magnetic susceptibility or other above-$T_c$ phenomena~\cite{Larkin-Book}. 

We extracted $T^*$ for each film by fixing field and sweeping temperature, as shown in Fig. \ref{Fig5}(d), resulting in the phase diagram shown in Fig. \ref{Fig1}(c). The existence of fluctuations far above the observed $T_c$ is expected for films approaching the critical thickness for superconductivity~\cite{Larkin-SIT}, however, surprisingly, the thickest films (100-300 nm) show positive magnetoresistance extending to $\sim$27 K (Fig. \ref{Fig5}(d)), significantly higher than the bulk transition $T_{c0}=11$ K. Clean, bulk superconductors are expected to have only a very narrow temperature regime where fluctuations manifest~\cite{Larkin-Book}. The superconducting coherence length of BPBO is $\sim$ 8 nm and isotropic, placing these thick films firmly in the 3D regime~\cite{Thanh,Batlogg}. Measurements on single crystals of BPBO reveal scaling consistent with a 2D material~\cite{GG1} and TEM measurements show some evidence of striped polymorph ordering on the scale of the coherence length~\cite{GG3}, suggesting nanostructuring of the material could contribute to this phenomenon. We, however, have no evidence of such stripes in our films, and our measurements suggest a high level of disorder in BPBO. The Mott-Ioffe-Regel parameter, $k_Fl$, extracted from room temperature measurements for the thick films is $\sim$3-4, consistent with bulk single crystal measurements~\cite{GG2,Batlogg}, and in the regime of a disordered material. This finding agrees with tunneling measurements on bulk BPBO crystals that show changes to the tunneling density of states arising from disorder~\cite{Luna}.

The role of disorder and superconducting fluctuations has interested the community, in part due to the layered structure of the high-$T_c$ cuprates and nearby insulating phases that promote the presence of fluctuations. Similarly, thin conventional superconductors exhibit strong fluctuations, proximity to a superconductor-insulator transition, and evidence for a pseudo-gap in the density of states, suggesting a connection between thin conventional superconductors and the higher-$T_c$ layered materials~\cite{Sacepe}. Disordered NbN films exhibit magnetoresistance inflection at temperatures close to the where the energy gap from scanning tunneling spectroscopy vanishes~\cite{Chand}. There is one report of a pseudogap in Ba$_{0.67}$K$_{0.33}$BiO$_3$~\cite{Chainani} and a pseudogap has been suggested for BPBO~\cite{Kitazawa}, however, experimentally the normal state properties are poorly explored in BBO superconductors.

Although the physical origin of disorder in BPBO is as of yet undetermined, there are several possibilities. Optical and terahertz spectroscopy measurements hint at local CDW fluctuations that could compete with superconductivity prior to the onset of a semiconducting bandgap from a long-range CDW~\cite{Tajima,Nicoletti}. In BPBO there is also evidence that many samples consist of two structural polymorphs, tetragonal and orthorhombic, and that superconductivity is correlated with the tetragonal volume fraction~\cite{Climent-Pascual,Marx}. Our measured $T^*$ of $\sim$27 K in thick films is very close to the higher $T_c$ of Ba$_{1-x}$K$_x$BiO$_3$, and in reasonable agreement with the prediction for disorder-free BPBO from Luna \textit{et al.}~\cite{Luna}. The similar temperature scales of our $T^*$, the disorder free $T_c$ prediction, and the onset of CDW fluctuations strongly suggests a connection of these phenomenon. Models of $T_c$ suppression, such as the Finkel'stein model, do not consider effects such as a competing CDW phase. The surprisingly large temperature range with measurable fluctuation effects are consistent with BPBO possessing a significantly higher $T_c$ that is obscured by disorder. 

Our work demonstrates a smaller critical thickness for films grown on LLO, however the lattice mismatch is still large compared to other perovskite heterostructures, preventing growth of films thick enough to study epitaxial strain engineering~\cite{Hwang,Chakhalian}. Further improvements in crystalline and interface qualities are important for BBO heterostructures that could use thin layers to control the CDW ~\cite{Kim} or interface superconducting BPBO with the predicted topological insulating phase of BBO~\cite{Yan}. Our results show how extrinsic contributions related to material limitations play an important role in the ultra-thin limit, as well as reveal new above-$T_c$ phenomenon in BPBO.

\subsection*{Acknowledgements}
 
Synthesis, characterization, and analysis were supported with funding from the Department of Energy Office of Basic Energy Sciences under award number DE-FG02-06ER46327. The theoretical input of A.L. was supported by NSF CAREER Grant No. DMR-1653661. We thank Art Hebard for helpful discussions and Jim Langyel for assistance with x-ray diffraction.
 
 \subsection*{Supplementary Information: Materials and Methods}
 
 \textit{Target preparation}. Targets of Ba(Pb$_{1-x}$Bi$_x$)$_y$O$_3$ were prepared from starting powders of Ba(NO$_3$)$_2$, PbO, and Bi$_2$O$_3$, with excess Pb and Bi in order to compensate for the relatively high vapor pressures of lead and bismuth oxides during sintering and growth. The starting powders were mixed in a planetary mill with zirconia media and isopropyl alcohol then dried overnight. Calcination was performed at 725$^\circ$ C for 24 hours in air followed by another planetary milling step. The resulting powder appeared to be single-phase perovskite within the limits of laboratory XRD. 2'' sputtering targets were formed in a uniaxial press. The green targets were covered top and bottom with excess BPBO powder to limit Pb and Bi volatility then sintered at 850$^\circ$ C for 1 hour in air. Excess Bi/Pb, $y$, was varied from 1 to 1.25, with films $y=1.08$ and below exhibiting poor stability and Bi/Pb deficiency as measured by RBS. Films with $y=1.20$ and above showed depressed superconducting transition temperatures. We therefore selected $y=1.13$.
  
\begin{figure}
		\centering
		\includegraphics[width=16cm]{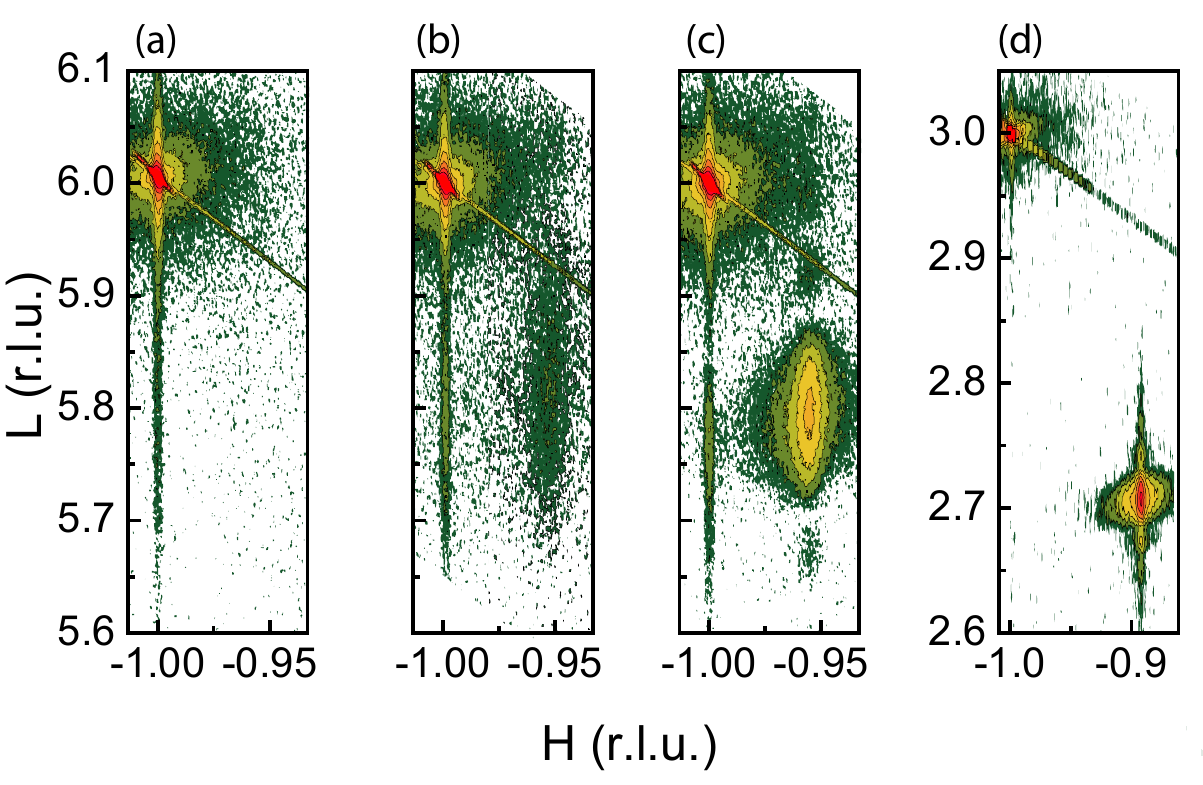}
		\caption{Reciprocal space maps of BaBiO$_3$. Reciprocal space maps around the 103$_{pc}$ reflection for (a) a fully coherent 4.6 nm thick film on LLO, as well bilayer relaxed films on LLO that are (b) 6.1 nm thick and (c) 12.1 nm thick. A (d) fully relaxed 18.3 nm thick film on STO. }\label{Fig6}
\end{figure}
  
 \textit{Film growth}. STO substrates (CrysTec GmbH) with single termination were prepared using buffered-HF etches and oxygen anneals. LLO (CrysTec GmbH) substrates were cleaned with acetone, methanol, and isopropyl alcohol, then annealed at 900$^\circ$ C in flowing oxygen for 2 hours, revealing a smooth step-terrace structure with step heights of $\sim$4.2 \AA (see Fig. \ref{Fig4}(b)). Longer and/or higher temperature annealing resulted in the formation of large surface particles, likely resulting from a small fraction of Lu occupying the A-site creating a slightly off-stoichiometric crystal. Films were grown using 90$^\circ$ off-axis RF-magnetron sputtering at substrate temperatures of 525$^\circ$ C in a 200 mTorr, 38:2 Ar:O$_2$ atmosphere with a $\sim$2.4 nm/min growth rate and subsequently cooled to room temperature in 450 Torr oxygen. Higher oxygen partial pressure during growth suppressed the transition temperature.
  
\begin{figure}
		\centering
		\includegraphics[width=16cm]{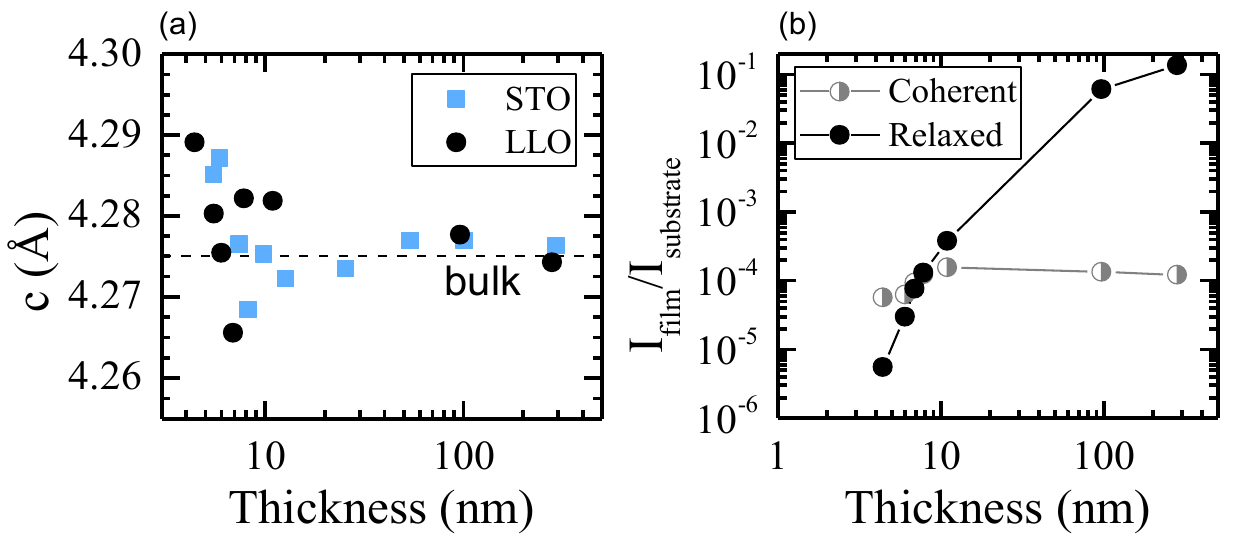}
		\caption{Extracted structural data from x-ray diffraction. (a) Lattice parameter vs. thickness for films on LLO and STO. The dashed line indicates the bulk $Ibmm$ value. (b) Relative intensity of coherent and relaxed phases on LLO.}\label{Fig7}
\end{figure}
  
 \textit{Film characterization}. X-ray diffraction was carried out on a Bruker D-8 in order to assess the crystalline quality. Out-of-plane scans used a scintillator point detector while reciprocal space maps were performed in asymmetric, grazing incidence mode with a line detector. Film thickness was measured using x-ray reflectivity. Surface topography was performed with a Bruker AFM. Electrical contacts were made using silver electrodes sputtered through a metal shadow mask in a square Van der Pauw configuration at the corners of the 5 mm $\times$ 5 mm sample. Magnetotransport measurements were performed in a Quantum Design Physical Properties Measurement System helium cryostat from 1.8 K to 300 K with fields up to 1.8 T.
 
\textit{Finkel'stein model for homogeneously disordered films}. For thin homogeneously disordered films, Finkel'stein developed a renormalization-group treatment for the critical temperature $T_c$ suppression from Coulomb interactions (hereafter $k_B=\hbar=1$)~\cite{Finkelstein}
\begin{equation}\label{F-Tc}
\frac{T_c}{T_{c0}}=\exp(-1/\gamma)\left[\left(1+\frac{\sqrt{t/2}}{\gamma-t/4}\right)\left(1-\frac{\sqrt{t/2}}{\gamma-t/4}\right)^{-1}\right]^{1/\sqrt{2t}}
\end{equation}
where $\gamma=1/\ln(T_{c0}\tau)$ is a disorder related fitting parameter that depends on the elastic scattering time $\tau$ and unrenormalized bulk transition temperature $T_{c0}$, while $t=e^2R_S/2\pi^2$. The dashed lines in Fig. \ref{Fig3}(c) represent fits to the initial low $R_S$ films and yield an increase in the scattering time $\tau$ by a factor of 3 for films on LLO compared to STO. 

As alluded in the main text, Eq. \eqref{F-Tc} does not accurately describe our data over the entire range of accessible parameters. This is not entirely surprising as the only parameter in Finkel'stein's model that controls the flow of $T_c$ is dimensionless conductance $t^{-1}$. There are various factors that can affect Eq. \eqref{F-Tc}. The importance of the exchange interaction was explored in a series of studies, most recently in Ref. ~\cite{Burmistrov}.  It was shown that the inclusion of a triplet channel coupling constant in the renormalization scheme can have a sizable effect on $T_c$. Furthermore, it was also emphasized that the effect of the Coulomb interaction on $T_c$ is sensitive to the cases of long-range versus short-range interaction models. In particular, in the latter case $T_c$ can be actually boosted to a higher value. More importantly for our experiments, Eq. \eqref{F-Tc} is expected to break down once film thickness $d$ becomes comparable to or greater than the mean free path $d>l$. Indeed, the initial gradual shift of $T_c$ is given by
\begin{equation}\label{Tc-ln3}
\frac{T_c-T_{c0}}{T_{c0}}=-\frac{t}{6}\ln^3\left(\frac{1}{T_{c0}\tau}\right) 
\end{equation}
which follows from Eq. \eqref{F-Tc} upon expansion over $\sqrt{t}|\gamma^{-1}|\ll1$. However, as pointed out by Finkel'stein~\cite{Finkelstein} for films with $d>l$ the cutoff of renormalization-group flow will become the Thouless energy $E_T=D/d^2$ instead of $\tau^{-1}$. As a result, for thicker films one expects the initial shift of $T_c$ to be given by  
\begin{equation}\label{Tc-ln3-2}
\frac{T_c-T_{c0}}{T_{c0}}=-\frac{1}{4(k_Fd)(k_Fl)}\ln^3\left(\frac{E_T}{T_{c0}}\right)
\end{equation}
instead of Eq. \eqref{Tc-ln3}, where we also used proper expression for the dimensionless conductance of a 3D film $t^{-1}=(2/3)(k_Fd)(k_Fl)$. As compared to Eq. \eqref{Tc-ln3} this result yields much weaker $T_c$ suppression as the logarithm contains an extra smallness in a parameter $l/d\ll1$ as $\ln\frac{E_T}{T_{c0}}=\ln\frac{(l/d)^2}{T_{c0}\tau}$, which practically reduces the effect by an order of magnitude because of its cubic dependence in Eq. \eqref{Tc-ln3-2}. In this regime one may be interested in considering other possible corrections to $T_c$ which are beyond the renormalization group treatment. In particular, electron-electron scattering with large momentum transfer, $\sim k_F$, may be important as renormalization group analysis that gives Eq. \eqref{F-Tc} captures only processes with typical momentum transfer of order $l^{-1}$. We are unaware of a detailed theory in this regime, but an estimate of these localizing corrections to $T_c$ gives $[T_c-T_{c0}]/T_{c0}\propto-1/(k_Fl)$ which is independent of $k_Fd\gg1$ (still assumes $k_Fl>1$) and thus can dominate over Eq. \eqref{Tc-ln3-2} for thick films.    

\begin{figure}
		\centering
		\includegraphics[width=16cm]{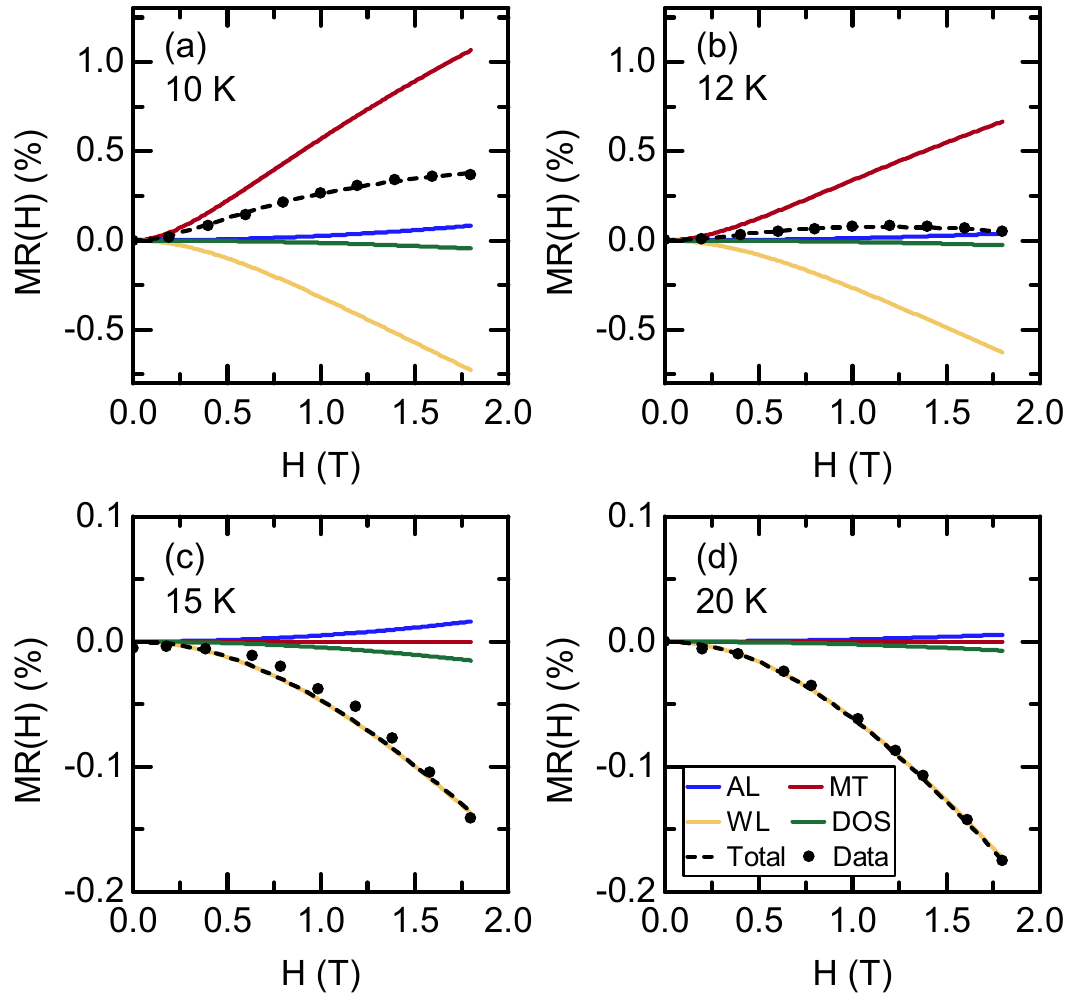}
		\caption{Fluctuation-induced and weak-localization fits to the 8.2 nm BPBO on STO sample shown in Fig. \ref{Fig5}(b). Field sweeps at (a) 10 K with fit parameters of $\tau_i =3.4\cdot10^{-12}$ s and $\tau_{WL} =1.6\cdot10^{-12}$ s; (b) 12 K with fit parameters of $\tau_i =2.8\cdot 10^{-12}$ s and $\tau_{WL} =1.5\cdot10^{-12}$ s; (c) 15 K with fit parameters of $\tau_i =1.0\cdot10^{-14}$ s and $\tau_{WL} =5.5\cdot10^{-13}$ s; and (d) 20 K with fit parameters of $\tau_i =1.0\cdot10^{-15}$ s and $\tau_{WL} =6.3\cdot10^{-13}$ s.}\label{Fig8}
\end{figure}
       
\textit{Fluctuations and magnetoresistance}. Gaussian fluctuations of Cooper pairs above $T_c$ are described by Aslamazov-Larkin (AL) and lead to a decrease in the resistance due to fluctuating short-circuits. The density-of-state (DOS) term arises from the reduction in the one-particle DOS when pairing occurs and increases the normal electron resistivity. The coherent scattering of pairs leads to the Maki-Thompson (MT) interference contribution. Finally, weak localization (WL) can contribute to the observed magnetoresistance (MR) and takes a similar form as the MT contributions. The total MR is then the sum of these terms. For thin disordered superconductors the fluctuation magnetoresistance contributions away from $T_c$ and at relatively weak fields are as follows \cite{Larkin-Book,Bergmann,Gordon,Shinozaki,Giannouri,Varlamov,Schwiete,AL,Rosenbaum}:
\begin{align}
&\frac{\delta R^{AL}_S(H,T)}{R^2_S}=-\frac{\pi G_Q}{\ln(T/T_c)}\left[\left(\frac{H_T}{H}\right)^2\left[\psi\left(\frac{1}{2}+\frac{H_T}{H}\right)-\psi\left(1+\frac{H_T}{H}\right)+\frac{H}{2H_T}\right]-\frac{1}{8}\right], \\ 
&\frac{\delta R^{MR}_S(H,T)}{R^2_S}=G_QA_{a}\left[\ln\left(\frac{H}{H_a}\right)+\psi\left(\frac{1}{2}+\frac{H_a}{H}\right)\right], \quad a=\{\mathrm{MT,DOS,WL}\}
\end{align}
where $G_Q=e^2/(2\pi)$, $H_T=(2T/\pi eD)\ln(T/T_c)$, and $\psi(x)$ is the digamma function. In the case of MR from MT, DOS, and WL contributions they all have the same functional form as governed by the function $\ln(x)+\psi(1/2+1/x)$ albeit with different amplitude factor $A_a$, and are sensitive to a different crossover field $H_a$. For the MT and WL cases the crossover field is determined by the dephasing processes, $H_{\mathrm{MT,WL}}=(4eD\tau_\phi)^{-1}$, where $\tau_\phi$ is the electron phase-breaking time, which is in general different for WL and MT contributions, see discussion in Ref.~[\onlinecite{Larkin-Book}] and references therein. For WL the Coulomb interaction leads to $\tau^{-1}_{\phi}\simeq T(R_SG_Q)\ln(1/R_SG_Q)$, however, interactions in the Cooper channel are also important. Very close to $T_c$ the Cooper channel gives a similar contribution to $\tau_\phi$, without a logarithmic factor, but with the bigger numerical pre-factor, while further away from $T_c$ dephasing from electron-fluctuation interactions is suppressed $\tau^{-1}_\phi\simeq T(R_SG_Q)\ln(1/T\tau)/\ln^2(T/T_c)$. Regularization of MT contribution is a subtler problem, which is dominated by the interactions in the Cooper channel and can be roughly estimated by $\tau^{-1}_\phi\simeq T\sqrt{R_SG_Q}$, which is parametrically distinct from the dephasing time of the WL term. Finally, in the case of the DOS term, the crossover field is $H_{\mathrm{DOS}}=H_T$. The amplitude factors are as follows: $A_{\mathrm{MT}}=\frac{1}{\pi}\beta(T/T_c)$, where $\beta$ is the Larkin's electron-electron interaction strength parameter in the Cooper channel. This function is tabulated and has following limiting behavior: $\beta=\pi^2/(6\ln^2(T/T_c))$ for $\ln(T/T_c)\gg1$  and  $\beta=\pi^2/(4\ln(T/T_c))$ for $\ln(T/T_c)\ll1$. The WL and DOS terms in MR have an opposite sign, namely $A_{\mathrm{WL}}=-\alpha_{\mathrm{SO}}/\pi$, where $\alpha_{\mathrm{SO}}$ is the dimensionless parameter of the strength of spin-orbit interaction, and $A_{\mathrm{DOS}}=-28\zeta(3)/\pi^3$.    
The diffusion coefficient $D$ is calculated from the slope of the upper critical field $D=-(4/\pi e)(dH_{c2}/dT)^{-1}_{T_c}$ and the transition temperature $T_c$ is taken to be the 50\% of the normal state value in resistivity measurements. This leaves the scattering times, $\tau_\phi$, and spin-orbit interaction constant, $\alpha_{\mathrm{SO}}$, as the fitting parameters. For all films $\alpha_{\mathrm{SO}}$ fits to the limiting case of $\alpha_{\mathrm{SO}}=1$, consistent with weak localization and small spin-orbit interaction with scattering impurities. Fig. \ref{Fig8} shows example fits for an 8.2 nm BPBO film on STO.


\end{document}